\documentclass[12pt]{article}
\makeatletter

\@addtoreset{equation}{section}
\makeatother

\newcommand{\nn}{\nonumber\\}
\newcommand{\pa}{\partial}
\def\bra{\langle}
\def\ket{\rangle}

\newcommand{\co}{\varphi^c}
\newcommand{\si}{\varphi^s}
\newcommand{\vs}[1]{\vspace{#1 mm}}

\newcommand{\e}{\epsilon}
\newcommand{\XB}{X_B}

\newcommand{\XD}{X_D}

\newcommand{\HD}{H_D^\epsilon}

\begin{document}

\rightline{\hbox{hep-th/0202200}}
\baselineskip 0.8cm
\vspace*{3.5cm}
\begin{center}
{\Large \bf Non-commutative Space And Chan-Paton Algebra}
\vskip 3mm
{\Large \bf In Open String Field Algebra}
\end{center}
\vskip 6ex
\begin{center}
{\large \sc FURUUCHI\ \ Kazuyuki} ${}^{1,2}$
\vskip 7mm

${}^1$ {\sl Department of Physics and Astronomy,}\\
${}^2$ {\sl Pacific Institute for the Mathematical Sciences,}\\
{\sl University of British Columbia, 
Vancouver, British Columbia, Canada V6T 1Z1}\\
\vskip 1mm
  {\tt furuuchi@physics.ubc.ca}
\end{center}
\vskip 5mm
%
\baselineskip=6mm
\centerline{{\bf Abstract}}
\vskip 4mm
There are several equivalent descriptions for
constant B-field background of open string.
The background can be 
interpreted as 
constant B-field as well as
constant gauge field strength or
infinitely many D-branes
with non-commuting Chan-Paton matrices.
In this article, 
the equivalence of 
these open string theories
is studied 
in Witten's cubic open string field theory.
Through the map between these 
equivalent descriptions,
both algebra of
non-commutative coordinates 
as well as Chan-Paton matrix algebra
are identified with subalgebras of 
open string field algebra.

\newpage
\tableofcontents
\section{Introduction}

Dirichlet branes (D-branes)
\cite{Dbranes}
have played crucial roles
in the study of non-perturbative
aspects of string theory.
D-branes are often referred to as solitons
in string theory.
However, to treat them really as solitons,
one needs to work in 
the framework of string field theory.
In open string field theory,
D-brane exists by definition because
open string with Neumann boundary conditions
in all space-time directions 
is equivalent to the open string
with its ends on a space-time filling D-brane.
Still, it is important to know
whether lower dimensional D-branes
are contained in open string field theory as solitons 
\cite{descent,TcondSFT}.

Open string theory
in constant B-field background
has a description in terms of 
non-commutative coordinates 
\cite{Connes,Hull,Schomerus,NCSW}.
Instantons and solitons \cite{Nikita,GMS} on
such non-commutative spaces
have been studied extensively and
interesting non-abelian 
structures similar to
Chan-Paton matrix algebra were found.
Indeed, these non-abelian structures were
regarded as low energy descriptions
of Chan-Paton matrix \cite{NCsolst},
and 
projection operators
\cite{FK1,GMS} or
partial isometries
\cite{FK2,solgene}
appeared as typical elements in the 
description of these instantons or solitons.
On the other hand,
Witten's cubic open string field theory itself 
defines a
non-commutative geometry \cite{NCSFT}.
The 3-string interaction vertex defines 
a non-commutative associative
product in open string field algebra.
It is tempting to expect that
microscopic description of D-branes
can be obtained by applying the techniques 
in non-commutative solitons
to Witten's cubic open string field theory.
Some
projection operators
in open string field algebra
have been constructed 
\cite{RZ,KP,RSZ,Matsuo1,Matsuo2,butt,GRSZ}
along this line of thought.

There are several equivalent 
descriptions for
open string in constant background B-field.
The background can be regarded as
constant B-field,
as well as constant gauge field strength
or infinitely many lower dimensional 
D-branes with
non-commuting Chan-Paton matrices.
In this article, I will call these descriptions as
B-picture, F-picture, and D-picture,
respectively.
The equivalence is most explicitly shown
by the method introduced by Ishibashi \cite{Ish}.
These descriptions are mapped to each other
by a change of variables.

In this article, I apply
the Ishibashi's method to 
study the equivalence
of these three descriptions in
Witten's cubic open string field theory.
This is a natural direction to pursuit
since field theory is an appropriate 
framework for the description of
backgrounds and change of variables. 

The organization of this article is as follows.
In section \ref{secBF},
I study the equivalence of
B-picture and F-picture.
Open string field theory
in B-picture was also studied
in \cite{Sgi,KT1,KT2}, where
it was found that 
the target space coordinates
become non-commutative.
It was also found 
that there is a map 
to the theory with commutative coordinates.
I point out this map is essentially the map
between B-picture and F-picture.
Although important steps have already been made in
these preceding works \cite{Sgi,KT1,KT2}, I believe
this new viewpoint makes the physical picture
of the map clearer.
Since F-picture variables
are the original variables, i.e.
the same to the variables without background
B-field,
this means that
the algebra of non-commutative
coordinates
is already contained
in the original open string field algebra.
I also give care to the treatment of subtlety in
the map
that arises at the boundaries of open string.
In section \ref{secD},
I study the map between
B-picture and D-picture.
From this map I obtain
the expression of
Chan-Paton matrices 
in terms of B-picture variables.
Through the map between B-picture and F-picture,
it is also possible to express them
in terms of F-picture variables.
This means that
Chan-Paton matrix algebra
is contained in the original
open string field algebra.
Since Chan-Paton matrix
is associated with lower dimensional D-branes, 
the appearance of the Chan-Paton algebra
in open string field algebra ensures
that lower dimensional D-branes can be described as some
field configurations 
in Witten's cubic open string field theory.
Section \ref{summary} is devoted to the summary and discussions.

\section{Equivalence Of B-Field And
Gauge Field Strength
In Open String Field Theory}\label{secBF}

\subsection{Review Of Witten's Open String Field Theory
In Background B-field}

\subsubsection*{Witten's Cubic Open String Field Theory}

I start from recalling the open string theory
without B-field background.
The worldsheet action is given by
\begin{eqnarray}
 \label{WS}
S=
-\frac{1}{4\pi\alpha'}
\int d\tau \int_0^{\pi} d\sigma
\left\{
g_{ij} \eta^{\alpha\beta} \pa_\alpha X^i \pa_\beta X^j
\right\} .
\end{eqnarray}
Here, 
$\eta^{\alpha\beta} \equiv \mbox{diag}(-1,1)\,\, 
(\alpha = 0,1)$,
$\sigma^0=\tau$,  $\sigma^1=\sigma$, and
the target space metric 
$g_{ij}\,\, (i,j = 0,\cdots, 25)$ is flat.
The equation of motion for $X^i(\tau,\sigma)$
is given by
\begin{eqnarray}
g_{ij}
\left(
\ddot{X}^j(\tau,\sigma) - {X^j}''(\tau,\sigma) 
\right) = 0,
\end{eqnarray}
where the differentials $\pa_\tau$ and $\pa_\sigma$
are denoted by dot $\dot{\,\,}$ and prime $\,'\,$, respectively.
There are
two possible boundary conditions.
One is the
Neumann boundary condition
\begin{eqnarray}
 \label{Nbc}
{X^i}'(\tau,\sigma) = 0, \quad (\sigma = 0, \pi)
\end{eqnarray}
and another is the 
Dirichlet boundary condition
\begin{eqnarray}
 \label{Dbc}
\dot{X}^i(\tau,\sigma) = const. \quad (\sigma = 0, \pi)
\end{eqnarray}
Here, I consider the case  
with Neumann boundary conditions
for all directions.
Then, $X^i(\tau,\sigma)$ can be expanded as
\begin{eqnarray}
 \label{Xtsmod}
X^i(\tau,\sigma) =
\sum_{n=0}^{\infty}
x^i_n(\tau) \co_n(\sigma),
\end{eqnarray}
where $\co_n(\sigma)$ are ortho-normal complete basis for
functions on $0\le \sigma \le \pi$
with Neumann boundary conditions at $\sigma = 0,\pi$:
\begin{eqnarray}
 \label{defco}
&&\co_n(\sigma) = \sqrt{\frac{2}{\pi}} \cos n \sigma,
 \quad (n \ne 0) \nn
&&\co_0(\sigma) = \sqrt{\frac{1}{\pi}} \quad.
\end{eqnarray}
I will use Hamiltonian formalism on the worldsheet
in order to describe string field theory
in operator formalism \cite{GJ,Samuel,Cremmer,NOhta}.
In Heisenberg picture, I have\footnote{
I use the same letters for the fields in
Lagrangian formalism
and Hamiltonian formalism.
Readers might not be confused in the context.
}
\begin{eqnarray}
 \label{NoBX}
X^i(\sigma) =
\sum_{n=0}^{\infty}
x^i_n \co_n(\sigma).
\end{eqnarray}
The canonical conjugate momentum $P_i(\sigma)$
of $X^i(\sigma)$
is given by
\begin{eqnarray}
 \label{NoBP}
P_i(\sigma)
= \frac{1}{2\pi\alpha'} g_{ij} \dot{X}^j(\sigma).
\end{eqnarray}
$P_i(\sigma)$ also 
satisfies Neumann boundary conditions and 
can be expanded as
\begin{eqnarray}
 \label{NoBPmod}
P_i(\sigma)
=\sum_{n=0}^{\infty} p_{in} \co_n (\sigma).
\end{eqnarray}
The canonical commutation relations
between modes 
turn out to be
\begin{eqnarray}
 \label{comcom}
[x_m^i,x_n^j]=0, \quad [p_{im},p_{j n}] = 0, \quad 
[x_m^i,p_{j n}] = i \delta^i_j \delta_{mn} .
\end{eqnarray}
The Hamiltonian of this system becomes
\begin{eqnarray}
 \label{Hmode}
H
&=&
\frac{1}{4\pi\alpha'}
\sum_{n=0}^\infty
\left\{
(2\pi\alpha')^2g^{ij} 
p_{in}p_{jn}
+
g_{ij} x^i_n x^j_n n^2
\right\} .
\end{eqnarray}
The equations of motion in Heisenberg picture are given by
\begin{eqnarray}
\dot{x}_n^i &=& i [H, x_n^i] = 2\pi\alpha' g^{ij} p_{jn}, \\
\dot{p}_{in}&=& i [H,p_{in}] 
= - \frac{1}{2\pi\alpha'} g_{ij}n^2 x^j_n.
\end{eqnarray}
To fix 
reparametrization invariance,
one introduces ghost action $S^{gh}$:
\begin{eqnarray}
S^{gh}
=
\int d^2\sigma
\left(
b_{++} \partial_- c^+
+
b_{--} \partial_+ c^-
\right).
\end{eqnarray}
The ghost fields have mode expansions
\begin{eqnarray}
 \label{cmodes}
&&c^+ (\tau,\sigma) 
= \sum_{n=-\infty}^{\infty}
c_n e^{-in(\tau+\sigma)}, \quad
c^- (\tau,\sigma) 
= \sum_{n=-\infty}^{\infty}
c_n e^{-in(\tau-\sigma)}, \\
&&b_{++} (\tau,\sigma) =
\sum_{n=-\infty}^{\infty}
b_n e^{-in(\tau+\sigma)},\quad 
b_{--} (\tau,\sigma) =
\sum_{n=-\infty}^{\infty}
b_n e^{-in(\tau-\sigma)}. 
\end{eqnarray}
The kinetic part of string field action is 
constructed using the BRST charge
\begin{eqnarray}
 \label{Q}
Q = \sum_{n=-\infty}^{\infty}
: c_n \left(L_{-n}^X + \frac{1}{2} L_{-n}^{gh}\right) : \,\, .
\end{eqnarray}
Here, the matter Virasoro generator $L_n^X$
is given by
\begin{eqnarray}
 \label{LX}
L_n^X = 
\frac{1}{2} g_{ij} \sum_{\ell=0}^\infty 
:\alpha_{n-\ell}^i \alpha_\ell^j : \,\, ,
\end{eqnarray}
where
\begin{eqnarray}
 \label{alphas}
\alpha_n^i = 
\frac{1}{\sqrt{4\pi\alpha'}}
(2\pi\alpha' g^{ij} p_{jn} - i n x_n^i) .
\end{eqnarray}
and $:\, \,:$ denotes the normal ordering.
The ghost Virasoro generator
$L_n^{gh}$
is given by
\begin{eqnarray}
 \label{Lgh}
L^{gh}_n
=
\sum_{m=-\infty}^{\infty}
(n-m): b_{n+m} c_{-m}: \,\,.
\end{eqnarray}
%
The action of the cubic
open string field theory is given
as follows \cite{NCSFT,GJ,Samuel,Cremmer,NOhta}
\begin{eqnarray}
 \label{cubic}
{\bf S} = 
\frac{1}{2} 
\left\langle R(1,2)\right| 
| \Psi \rangle_1
\left(
Q^{(2)} | \Psi \rangle_2
\right)
+\frac{g}{3}\, \,  {}_{123}\left\langle V_3\right|  
| \Psi \rangle_1
| \Psi \rangle_2
| \Psi \rangle_3  .
\end{eqnarray}
Here,
the 3-string interaction vertex is defined by 
the overlap condition
\begin{eqnarray}
 \label{overlap}
&& \bra V_3 |
\left( X^{i(r+1)}(\sigma) - X^{i (r)}(\pi-\sigma) \right) = 0, 
\quad (0 \le \sigma \le \frac{\pi}{2} )   \nn
&& \bra V_3 |
\left(P_i^{(r+1)}(\sigma) + P^{(r)}_i(\pi-\sigma)\right) = 0.
\qquad (0 \le \sigma \le \frac{\pi}{2} )
\end{eqnarray}
The matter part of the string vertex has a form
\begin{eqnarray}
 \label{V3}
\bra V_3^X |
=
{}_{123} \bra 0 | 
\exp
\left(-\frac{1}{2}
g_{ij}\sum_{\stackrel{r,s=1,2,3}{m,n\geq 0}} 
a^{i(r)}_m V_{3\,mn}^{rs} a_n^{j(s)} \right),
\end{eqnarray}
where creation operators $a_n^{i \dagger}$
and annihilation operators
$a_n^i$  are defined by
\begin{eqnarray}
 \label{cre}
a_n^i &=& \frac{1}{\sqrt{4\pi\alpha'n}}
(2\pi\alpha' g^{ij}p_{jn} - i n x_n^i), 
\quad (n \ge 1)  \nn
a_n^{i \dagger} &=& \frac{1}{\sqrt{4\pi\alpha'n}}
(2\pi\alpha' g^{ij}p_{jn} + i n x_n^i),
\quad (n \ge 1) \nn
a_0^i &=& 
\frac{1}{\sqrt{4\pi\alpha'}}
(4\pi\alpha' g^{ij}p_{j0} - i x_0^i), \nn
a_0^{i \dagger} &=& 
\frac{1}{\sqrt{4\pi \alpha'}}
(4\pi\alpha' g^{ij} p_{j0} + i x_0^i).
\end{eqnarray}
In this article, the explicit
form of the coefficients $V_{3\,mn}^{rs}$
is not used,
see \cite{GJ,Samuel,Cremmer,NOhta} for details. 
Also
I do not study
the ghost part since it is
not modified by constant background B-field.
The matter part of the 
identity string field $|I\ket $, 
the reflector $\bra R(1,2) |$ 
and the inverse reflector $|R(1,2)\ket$
are given by
\begin{eqnarray}
 \label{Istate}
|I\ket &=& 
\exp
\left(
-\frac{1}{2} g_{ij} 
\sum_{n,m}
a^{i\dagger}_n C_{nm} a_m^{j\dagger}
\right)
|0\ket, \\
\bra R(1,2)| &=&
{}_{12}\bra 0 | \exp
\left(
-g_{ij} 
\sum_{n,m}
a^{i(1)}_n C_{nm} a_m^{j(2)}
\right), \label{Ref}\\
| R(1,2) \ket
&=&
\exp
\left(
-g_{ij} 
\sum_{n,m}
a^{i \dagger (1)}_n C_{nm} a_m^{j \dagger (2)}
\right) | 0 \ket_{12},
\label{IRef}
\end{eqnarray}
where $C_{nm}=(-1)^n\delta_{nm}$.
$| I \ket$ and $\bra R(1,2)|$
satisfy the following overlap conditions
\begin{eqnarray}
 \label{overlapId}
\left(X^i(\sigma) - X^i(\pi-\sigma)\right)| I \ket &=& 0, \\
\left(P_i(\sigma) + P_i(\pi-\sigma)\right)| I \ket &=& 0,
\end{eqnarray}
\begin{eqnarray}
  \label{overlapR}
&& \bra R(1,2)| \left(X^{i(r+1)}(\sigma) - X^{i(r)}(\pi-\sigma)\right) = 0, 
\quad (0 \le \sigma \le \frac{\pi}{2})\\
&& \bra R(1,2)| \left(P_i^{(r+1)}(\sigma) + P_i^{(r)}(\pi-\sigma)\right) = 0.
\quad \,\,\,\,\, (0 \le \sigma \le \frac{\pi}{2})
\end{eqnarray}
The star product of string fields is defined by
\begin{equation}
 \label{star}
 |A\star B\ket_1=
 {}_{234}\bra V_3 |
A \ket_3 |B \ket_4 | R (1,2) \ket.
\end{equation}
One can define integration 
$\int$
in string field algebra by
\begin{eqnarray}
 \int \Psi = {}_1\langle I | \Psi \rangle_1,\quad
 {}_1\bra I | =
 \bra R(1,2) | I \ket_{2}.
\end{eqnarray}
The identity string field satisfies\footnote{%
I dropped the normalization factor for simplisity.
}
\begin{eqnarray}
 \label{identity}
|I\star  I \ket = | I \ket.
\end{eqnarray}
The open string 
field action (\ref{cubic}) can be rewritten 
into the following form 
\begin{eqnarray}
 \label{NCcubic}
\mbox{\bf S} = \int
\frac{1}{2} \Psi \star (Q \Psi)
+\frac{g}{3}  \Psi \star \Psi \star \Psi.
\end{eqnarray}

\subsubsection*{Open String Theory In Constant %
Background B-field}

Next, I review the open string theory 
in constant background B-field $B_{ij}$.
This background is described by the worldsheet action
\begin{eqnarray}
 \label{BWS}
S=
-\frac{1}{4\pi\alpha'}
\int d\tau \int_0^{\pi} d\sigma
\left\{
g_{ij} \eta^{\alpha\beta} \pa_\alpha X^i \pa_\beta X^j
-
2\pi\alpha' \epsilon^{\alpha\beta}
B_{ij}\pa_\alpha X^i \pa_\beta X^j
\right\} .
\end{eqnarray}
Here,
$\epsilon^{01} = 1$.
To avoid the special features of space-time
non-commutativity \cite{SST},
I only consider the case $B_{0i}\ne0$.
The equations of motion is given by
\begin{eqnarray}
  \label{eqXB}
g_{ij}
\left(
\ddot{\XB}^j(\tau,\sigma)-{\XB^j}''(\tau,\sigma)
\right) = 0,
\end{eqnarray}
where I have introduced the suffix $B$ to 
explicitely 
indicate the dependence on the B-field.
The boundary conditions are modified from (\ref{Nbc}) by
the presence of the B-field:
\begin{eqnarray}
 \label{bcond}
g_{ij}{\XB^{j}}'(\tau,\sigma)+b_{ij}\dot{\XB}^j(\tau,\sigma) = 0.
\quad (\sigma = 0, \pi) 
\end{eqnarray}
Here,
$b_{ij} = 2\pi\alpha' B_{ij}$.
In Hamiltonian formalism,
the boundary conditions (\ref{bcond}) can be
treated as constraints \cite{ChuHo}
(I use notation similar to \cite{KT1}).
The mode expansions of $\XB^i(\sigma)$
and its canonical variable $\hat{P}_i(\sigma)$
are given as follows:
\begin{eqnarray}
 \label{XhX}
\XB^i(\sigma) = \hat{X}^{i}(\sigma) 
+ \theta^{ij} Q_j(\sigma),
\end{eqnarray}
\begin{eqnarray}
 \label{hP}
\hat{P}_i(\sigma)
=
\frac{1}{2\pi\alpha'}
\left\{
g_{ij} \dot{\XB}^j(\sigma) + b_{ij} {\XB^j}'(\sigma)
\right\}, 
\end{eqnarray}
\begin{eqnarray}
 \label{hXQ}
\hat{X}^i(\sigma)  =
\sum_{n=0}^{\infty} \hat{x}^i_n \co_n(\sigma),
\quad
Q_i(\sigma)  = 
\hat{p}_{i0} \sqrt{\frac{1}{\pi}}\left(\sigma - \frac{\pi}{2} \right) +
\sum_{n=1}^{\infty} \frac{1}{n} \hat{p}_{i n} \si_n(\sigma),
\end{eqnarray}
\begin{eqnarray}
 \label{hPmod}
\hat{P}_i(\sigma)
=
\sum_{n=0}^{\infty} \hat{p}_{i n} \co_n (\sigma) .
\end{eqnarray}
Here, $\si_n(\sigma)$ are ortho-normal
basis for functions on $0\le \sigma \le \pi$
with Dirichlet boundary conditions
at $\sigma = 0,\pi$:
\begin{eqnarray}
 \label{defsi}
\si_n(\sigma) = \sqrt{\frac{2}{\pi}} \sin n \sigma ,
\end{eqnarray}
and $\theta^{ij}$ is defined by
\begin{eqnarray}
 \label{theta}
\frac{\theta^{ij}}{2\pi \alpha'}
&=& -  
\left(
\frac{1}{g+b} b \frac{1}{g-b}
\right)\!{}^{ij}.
\end{eqnarray}
The modes $\hat{x}^i_m$ and  $\hat{p}_{j n}$
satisfy the  canonical commutation relations
\begin{eqnarray}
 \label{commB}
[\hat{x}^i_m, \hat{x}^j_n] = 
[\hat{p}_{i m},\hat{p}_{j n}] = 0, \quad
[\hat{x}^i_m, \hat{p}_{j n} ] = i \delta_{mn} \delta^i_j.
\end{eqnarray}
These commutation relations
lead to the following commutation relations
\begin{eqnarray}
 \label{commXP}
&&[ \XB^i(\sigma), \XB^j(\sigma')] 
=
i \theta^{ij}
\left(
 \delta_{0,\sigma+\sigma'}
-\delta_{2\pi,\sigma+\sigma'}
\right), \nn
&&[ \XB^i(\sigma), \hat{P}_j(\sigma')] = 
i \delta^i_j \delta^c (\sigma, \sigma'),  \quad
[\hat{P}_{i}(\sigma),  \hat{P}_j(\sigma') ] = 0,
\end{eqnarray}
where
\begin{eqnarray}
  \label{Krod}
 \begin{array}{lll}
\delta_{0,\sigma} &=& 1, \quad (\sigma = 0) \\
                  &=& 0, \quad (\sigma \ne 0)
 \end{array}
\qquad
 \begin{array}{lll}
\delta_{\pi,\sigma} &=& 1, \quad (\sigma = \pi) \\
                    &=& 0, \quad (\sigma \ne \pi)
 \end{array}
\end{eqnarray}
and
\begin{eqnarray}
 \label{deltac}
\delta^c (\sigma,\sigma') =
\sum_{n=0}^{\infty} \co_n(\sigma) \co_n(\sigma').
\end{eqnarray}
Notice the non-commutativity of $\XB^i(\sigma)$
at the boundaries.
The Hamiltonian $\hat{H}(B)$ of this system
becomes
\begin{eqnarray}
\hat{H}(B) = \frac{1}{4\pi\alpha'}
\sum_{n=0}^\infty
\left\{
(2\pi\alpha')^2 G^{ij}\hat{p}_{i n} \hat{p}_{j n} 
+
G_{ij} \hat{x}^i_n \hat{x}^j_n n^2
\right\},
\end{eqnarray}
where
\begin{eqnarray}
 \label{G}
G_{ij} &=& g_{ij} - (b g^{-1} b)_{ij}, \nn
G^{ij} &=&
\left(
\frac{1}{g+b} g \frac{1}{g-b}
\right)\!{}^{ij}.
\end{eqnarray}
$B$ of $\hat{H}(B)$ indicates the dependence on 
the background B-field.
The equations of motion
in Heisenberg picture are given by
\begin{eqnarray}
\dot{\hat{x}}^i_n &=& 
i[\hat{H}(B), \hat{x}^i_n]=
2\pi\alpha' G^{ij} \hat{p}_{jn}, \label{eqbx}\\
\dot{\hat{p}}_{i n} &=& 
i[\hat{H}(B),\hat{p}_{in}]
=
- \frac{1}{2\pi\alpha'} G_{ij} \hat{x}^j_n n^2 . \label{eqbp}
\end{eqnarray}


\subsection{Map Between B-picture and F-picture}

In the previous subsection,
the effect of the B-field 
is treated as boundary conditions (\ref{bcond}).
Since
the second term in the action (\ref{BWS})
is a total derivative,
it can be integrated to the surface term:
\begin{eqnarray}
 \label{FWS}
S&=&
-\frac{1}{4\pi\alpha'}
\int d\tau \int_0^{\pi} d\sigma
\left\{
g_{ij} \eta^{\alpha\beta} \pa_\alpha X^i \pa_\beta X^j
-
2\pi\alpha' \epsilon^{\alpha\beta}
B_{ij}\pa_\alpha X^i \pa_\beta X^j \right\} \nn
&=&
-\frac{1}{4\pi\alpha'}
\int d\tau \int_0^{\pi} d\sigma
\left\{
g_{ij} \eta^{\alpha\beta} \pa_\alpha X^i \pa_\beta X^j 
\right\} \nn
&&
-
\frac{1}{4\pi\alpha'}
\int d\tau \, \,
b_{ij}
X^i (\pi) \dot{X}^j (\pi) 
+
\frac{1}{4\pi\alpha'}
\int d\tau \, \,
b_{ij}
X^i (0)\dot{X}^j (0) .
\end{eqnarray}
Then, the effect of $B_{ij}$ appears as 
interactions at the boundaries.
It should be 
regard as
open string background, i.e.
constant background
gauge field strength.
I call this description F-picture,
and call the description in the
previous subsection
B-picture.
These descriptions should be
equivalent since both start from
the same action.
The equivalence is most explicitly
shown by
the method introduced by Ishibashi \cite{Ish},
and I will utilize his method.
In order to clearly distinguish the 
boundary interactions
from the boundary conditions,
one first shifts the positions of the
boundary terms to
$\sigma = \e$ and $\sigma = \pi - \e$, $(\e > 0)$
and take the limit $\e \rightarrow 0$ later.
This procedure ensures that the boundary terms
contribute to the equation of motion,
instead of changing the boundary conditions.
Then, the equation of motion is given by
\begin{eqnarray}
 \label{eqXF}
 g_{ij}
\left(
\ddot{X^j}(\tau,\sigma) - {X^j}''(\tau,\sigma)
\right) 
+ b_{ij}
\left(
\dot{X}^j(\tau,\pi-\e) 
\delta^c(\pi-\e,\sigma)
-
\dot{X}^j(\tau,\e) 
\delta^c(\e,\sigma) 
\right) = 0 ,
\end{eqnarray}
with Neumann boundary conditions:
\begin{eqnarray}
 \label{Neu}
g_{ij} {X^j}'(\tau,\sigma) = 0. \quad (\sigma = 0,\pi)
\end{eqnarray}
$X^i(\tau, \sigma)$ can be expanded as
\begin{eqnarray}
 \label{Xmode}
X^i(\tau,\sigma) =
\sum_{n=0}^{\infty}
x^i_n(\tau) \co_n(\sigma).
\end{eqnarray}
In terms of the modes, 
the equations of motion (\ref{eqXF})
can be rewritten as
\begin{eqnarray}
 \label{eqXFmode}
g_{ij} (\ddot{x}_n^j+ n^2x_n^j) +
b_{ij} 
\left(
\dot{X}^j(\pi-\epsilon) \co_n(\pi-\epsilon)
-\dot{X}^j(\epsilon) \co_n(\epsilon)
\right) &=& 0.
\end{eqnarray}
The canonical momentum $P_i(\sigma)$
of $X^i(\sigma)$ is given by
\begin{eqnarray}
 \label{FP}
P_i(\sigma) 
= \frac{1}{2\pi\alpha'} 
\left(
g_{ij} \dot{X}^j(\sigma) 
+
\frac{1}{2}
b_{ij} 
  \left(
X^j(\pi-\e) \delta^c(\pi-\e,\sigma) 
-X^j(\e) \delta^c( \e ,\sigma) 
  \right)
\right).
\end{eqnarray}
$P_i(\sigma)$ also satisfies the Neumann boundary conditions
and can be expanded as
\begin{eqnarray}
 \label{Pmode}
P_i(\sigma) =
\sum_{n=0}^{\infty} p_{in} \co_n (\sigma).
\end{eqnarray}
The variables in F-picture,
$x_n^i$ and $p_{i n}$,
are identified with the variables
without the background B-field, i.e.
(\ref{NoBX}) and (\ref{NoBP}),
respectively.
What is changed by the B-field is
the Hamiltonian.
It is given by
\begin{eqnarray}
 \label{modeHI}
H_I^\e
&=&
\frac{1}{4\pi\alpha'}
\sum_{n=0}^\infty
g_{ij} 
\left(
\dot{x}^i_{n} \dot{x}^j_{n} + 
n^2 x^i_n x^j_n
\right),
\end{eqnarray}
where the suffix $\e$ indicates
the dependence on the cut off and
$\dot{x}_n^i$ is solved in terms of 
$p_{i n}$ and $x_n^i$:
\begin{eqnarray}
 \label{eqxn}
\dot{x}^i_n &=& 
2\pi\alpha' g^{ij} p_{jn}
- \frac{1}{2} {(g^{-1}b)^i}_j
\left(
X^j(\pi-\e)\co_n(\pi-\e)-X^j(\e)\co_n(\e)
\right).
\end{eqnarray}
This is also an
equation of motion of 
$x_n^i$
in Heisenberg picture. 
The equation of motion for $p_{i n}$ is given by
\begin{eqnarray}
 \label{eqpn}
\dot{p}_{in} &=& i [H_I^\e , p_{in}] \nn
&=&
-\frac{1}{2\pi\alpha'} 
g_{ij} n^2 x_n^j
-\frac{1}{2}
 b_{ij}
\left(
(\dot{X}^j(\pi-\e)\co_n(\pi-\e)-\dot{X}^j(\e)\co_n(\e)
\right) .
\end{eqnarray}
%
%
%
%
I will show the equivalence of the two 
descriptions by constructing a one-to-one map
between B-picture variables
$(\hat{x}^i_n, \hat{p}_{i n})$
and F-picture variables
$(x^i_n, p_{in})$. 
This map is obtained
in the following way \cite{Ish}.
In the interval
$0 \le \sigma \le \pi$,
one can Fourier expand 
$(\sigma-\frac{\pi}{2})$ and $\si_m(\sigma)$ 
by $\co_n (\sigma)$,
and so as $\XB^i(\sigma)$.
Then, one
equates $X^i(\sigma) = \XB^i(\sigma)$
in the interval $0 < \sigma < \pi$.
By comparing the coefficients of $\co_n (\sigma)$,
one obtains the expression of 
$x^i_n$ in terms of 
$\hat{x}^i_n$ and $\hat{p}_{in}$.
Here, the limit $\epsilon \rightarrow 0$ is taken 
as
\begin{eqnarray}
 \label{limitF}
&& \lim_{\e \rightarrow 0}
X^i (\pi-\e) = X^i(\pi) \ne \XB^i (\pi), \nn
&& \lim_{\e \rightarrow 0}
X^i (\e) = X^i(0) \ne \XB^i (0) .
\end{eqnarray}
This prescription can be understood
from the corresponding boundary state 
in closed string theory \cite{Ish}\cite{Wilson,OkuB}.
Then, one can use (\ref{eqpn})
to obtain the
expression of $p_{in}$ in terms of
$\hat{x}^i_n$ and $\hat{p}_{in}$.
The result is as follows:
\begin{eqnarray}
x_n^i &=& 
\hat{x}_n^i 
- \theta^{ij} \sum_{\ell=0}^\infty \hat{p}_{j \ell} A_{\ell n} ,
\label{x} \\
p_{i n}&=&
\hat{p}_{i n}
-\frac{1}{2\pi\alpha'} b_{ij}
\sum_{\ell = 0}^\infty 
\left(
\hat{x}^j_\ell 
- 
\theta^{jk} 
\sum_{m=0}^\infty \hat{p}_{k m} A_{m \ell} 
\right)
A_{\ell n} \frac{\ell^2+n^2}{2},
\label{p}
\end{eqnarray}
where the matrix $A_{mn}$ is defined as
\begin{eqnarray}
 \label{cosi}
&&\co_n(\sigma) = \sum_{m=1}^\infty A_{nm} m \si_m(\sigma), 
\quad (0<\sigma<\pi) \nn
&&\si_n(\sigma) = \sum_{m=0}^\infty (-n) A_{nm}  \co_m(\sigma) .
\quad (0 \leq \sigma \leq \pi)
\end{eqnarray}
Some properties of the matrix
$A_{mn}$ is summarized 
in the appendix.
Here, one
must be careful since the infinite summations
may depend on ordering.
In (\ref{p}),
the bracket $(\,\,)$ means that 
the summation over mode indices inside
$(\,\,)$
should be taken before the summation over mode indices 
in and outside the bracket.
As an illustration, let us consider
the matter fields at the 
``boundary'':
\begin{eqnarray}
 \label{boundary}
&&X^i (\pi) = 
\sum_{n=0}^\infty
x^i_n \co_n (\pi)
=
\sum_{n=0}^\infty
\left(\hat{x}^i_n - 
\theta^{ij} \sum_{\ell = 0}^\infty \hat{p}_{j\ell} A_{\ell n}
\right) \co_n (\pi) \\
&\ne&
\XB^i (\pi)
=
\sum_{n=0}^\infty
\hat{x}^i_n \co_n(\pi)- 
\theta^{ij} \sum_{\ell = 0}^\infty \hat{p}_{j\ell} 
\left(
\sum_{n=0}^\infty
A_{\ell n}
\co_n (\pi)
\right)
=
\sum_{n=0}^\infty
\hat{x}^i_n \co_n(\pi)
+
\frac{\sqrt{\pi}}{2} 
\theta^{ij}
\hat{p}_{j0}. \nonumber
\end{eqnarray}
One must regard 
$X^i(\pi)$ and $\XB^i(\pi)$ as
different operators
since 
while $[X^i(\pi), X^j(\pi)]=0$,
$[\XB^i(\pi), \XB^j(\pi)] = - i \theta^{ij}$.
The difference between $X^i(\pi)$ and $\XB^i(\pi)$
may be interpreted as the difference of regularization
prescription appeared in \cite{NCSW}.

By applying the rule (\ref{limitF}),
the $\e \rightarrow 0$ limit of the
equations of motion (\ref{eqxn}) and (\ref{eqpn})
become
\begin{eqnarray}
 \label{limeqxn}
\dot{x}^i_n &=& 
2\pi\alpha' g^{ij} p_{jn}
- \frac{1}{2}  {(g^{-1}b)^i}_j
\left(
X^j(\pi)\co_n(\pi)-X^j(0)\co_n(0)
\right), \\
 \label{limeqpn}
\dot{p}_{in} 
&=&
-\frac{1}{2\pi\alpha'} 
g_{ij} n^2 x_n^j
-\frac{1}{2}
 b_{ij} 
\left(
(\dot{X}^j(\pi)\co_n(\pi)-\dot{X}^j(0)\co_n(0)
\right) .
\end{eqnarray}
One can check that 
the equations of motion in
B-picture 
((\ref{eqbx}) and (\ref{eqbp}))
lead to the equations of
motion in F-picture 
((\ref{limeqxn}) and (\ref{limeqpn})).
Thus the same Hamiltonian $\hat{H}(B)$ 
gives the equations of motion
of both pictures.
Hence the two descriptions are equivalent.

The inverse map is given by
\begin{eqnarray}
 \hat{x}^i_n
&=&
{(G^{-1}g)^i}_j
x^j_n 
+
\theta^{ij} \sum_{\ell=0}^\infty  p_{j \ell} A_{\ell n}  
+ \delta_{n0} \frac{\sqrt{\pi}}{4}
{{(\theta b)}^i}_j
\left(
X^j(\pi) + X^j(0)
\right),
\label{hx}\\
\hat{p}_{i n} &=&
p_{in}
+ \frac{1}{2\pi\alpha'} b_{ij}
\sum_{\ell=0}^\infty x_\ell^j A_{\ell n} \frac{\ell^2+n^2}{2}.
\label{hp}
\end{eqnarray}
The
map can be written in the form of
Unitary transformation:
\begin{eqnarray}
  \label{Bogoh}
x_n^i = U_{\scriptscriptstyle BF}^{-1} 
\hat{x}_n^i U_{\scriptscriptstyle BF}, \quad 
p_{i n} = U^{-1}_{\scriptscriptstyle BF} 
\hat{p}_{i n} U_{\scriptscriptstyle BF}.
\end{eqnarray}
Here, $U_{\scriptscriptstyle BF}$ is given by
\begin{eqnarray}
 \label{BogoU}
U_{\scriptscriptstyle BF} = e^N e^M,
\end{eqnarray}
where
\begin{eqnarray}
 \label{defM}
M &=& 
-\frac{i}{2} \theta^{ij} 
\sum_{\ell=0}^\infty
\sum_{m=0}^\infty \hat{p}_{i \ell} \hat{p}_{j m} 
A_{\ell m} \nn
&=&
-\frac{i}{4} \theta^{ij} \int_0^\pi d\sigma 
\int_0^\pi d\sigma'
\e (\sigma-\sigma') \hat{P}_i(\sigma ) \hat{P}_j (\sigma' ), 
\end{eqnarray}
and
\begin{eqnarray}
 \label{defN}
N &=& 
-\frac{i}{4\pi \alpha'} b_{ij} 
\sum_{\ell=0}^\infty \sum_{m =0}^\infty
\hat{x}_{\ell}^i A_{\ell m} \frac{\ell^2 + m^2}{2} \hat{x}_m^j \nn
&=& 
-\frac{i}{4} \int_0^\pi d\sigma 
B_{ij} 
\left(
 \hat{X}^i(\sigma) \hat{X}^{j}{}' (\sigma) -
 \hat{X}^{i}{}' (\sigma) {\hat{X}^j}(\sigma)
\right).
\end{eqnarray}
Here, $\e (\sigma)$
is the sign function
\begin{eqnarray}
\e (\sigma) 
&=& 1 , \quad \,\,\,(\sigma > 0) \nn
&=& 0 , \quad \,\,\,(\sigma = 0) \nn
&=& -1 . \quad(\sigma <0)
\end{eqnarray}
These operators
have appeared
in \cite{Sgi,KT1,KT2}.

The 3-string vertex
$\bra V_3 |$
can be rewritten
using the variables $\hat{x}_n^i$ and $\hat{p}_{i n}$.
From (\ref{Bogoh}), I obtain
\begin{eqnarray}
\bra V_3 | = 
\bra \hat{V}_3 | e^N e^M ,
\end{eqnarray}
where $\bra \hat{V}_3 |$ satisfies
the following 
overlap conditions 
for $\hat{X}^i(\sigma)$ and $\hat{P}_i (\sigma)$:
\begin{eqnarray}
 \label{hoverlap}
&& \bra \hat{V}_3 |\left(\hat{X}^{i(r+1)}(\sigma) 
- \hat{X}^{i(r)}(\pi-\sigma)\right) = 0, 
\quad  ( 0 \le \sigma \le \frac{\pi}{2} )\nn
&& \bra \hat{V}_3 |\left(\hat{P}_i^{(r+1)}(\sigma) 
+ \hat{P}_i^{(r)}(\pi-\sigma)\right) = 0 .
\qquad  ( 0 \le \sigma \le \frac{\pi}{2} )
\end{eqnarray}
Since 
the difference between the original
overlap conditions (\ref{overlap})
and (\ref{hoverlap})
is just a difference of the 
variables,
\footnote{
Since the overlap conditions
do not refer to metric, the 3-string vertex is
independent from the metric \cite{KugoZ,KT1}.}
\begin{eqnarray}
 \label{hV3}
\bra \hat{V}_3^X |
=
{}_{123} \bra 0 | 
\exp
\left(
-\frac{1}{2} g_{ij}
\sum_{\stackrel{r,s=1,2,3}{m,n\geq 0}} 
\hat{a}^{i(r)}_m V_{3\,mn}^{rs} \hat{a}_n^{j(s)} 
\right),
\end{eqnarray}
where the 
creation and annihilation operators
are defined in the same way as in (\ref{cre}) :
\begin{eqnarray}
 \label{hcre}
\hat{a}_n^i &=& \frac{1}{\sqrt{4\pi\alpha'n}}
(2\pi\alpha' g^{ij} \hat{p}_{jn} - i n \hat{x}_n^i), 
\quad (n \ge 1) \nn
\hat{a}_n^{i \dagger} &=& \frac{1}{\sqrt{4\pi\alpha'n}}
(2\pi\alpha' g^{ij}  \hat{p}_{jn} + i n \hat{x}_n^i),
\quad (n \ge 1) \nn
\hat{a}_0^i &=& \frac{1}{\sqrt{4\pi\alpha'}}
(4\pi\alpha' g^{ij}\hat{p}_{j0} - i n \hat{x}_0^i), \nn
\hat{a}_0^{i \dagger} &=& \frac{1}{\sqrt{4\pi\alpha'}}
(4\pi\alpha' g^{ij}\hat{p}_{j0} + i n \hat{x}_0^i) .
\end{eqnarray}
Using the defining properties (\ref{hoverlap}),
one can show
$\bra \hat{V}_3 | N = 0$.
From the overlap conditions,
\begin{eqnarray}
 \label{NtoV3}
& &\bra \hat{V}_3 | 
\sum_{r=1}^3
\int_0^\pi d\sigma 
 \hat{X}^{i(r)}(\sigma) \hat{X}^{j(r)}{}' (\sigma) \nn
&=&
\bra \hat{V}_3 | 
\sum_{r=1}^3
-\left(
\int_0^{\frac{\pi}{2}} d\sigma 
 \hat{X}^{i(r)}(\pi-\sigma) 
 \hat{X}^{j(r)}{}' (\pi-\sigma) 
+
\int^\pi_{(\frac{\pi}{2})} d\sigma 
 \hat{X}^{i(r)}(\pi-\sigma) 
 \hat{X}^{j(r)}{}' (\pi-\sigma) 
\right)\nn
&=&
\bra \hat{V}_3 | 
\sum_{r=1}^3
\left(
\int_\pi^{\frac{\pi}{2}} d\sigma 
 \hat{X}^{i(r)}(\sigma) 
 \hat{X}^{j(r)}{}' (\sigma) 
+
\int^0_{(\frac{\pi}{2})} d\sigma 
 \hat{X}^{i(r)}(\sigma) 
 \hat{X}^{j(r)}{}' (\sigma)
\right) \nn
&=&
-
\bra \hat{V}_3 | 
\sum_{r=1}^3
\int_0^\pi d\sigma 
 \hat{X}^{i(r)}(\sigma) \hat{X}^{j(r)}{}' (\sigma) \nn
&=&0 ,
\end{eqnarray}
where $\int_{(p)}$
means that 
the integration region starts from
the point $p$ with the point $p$
itself being
excluded from the integration.
The equation (\ref{NtoV3})
means $\bra \hat{V}_3 | N = 0$.
Thus, I obtain
\begin{eqnarray}
 \label{VhV}
\bra V_3 | = 
\bra \hat{V}_3 | e^M .
\end{eqnarray}
This expression was also found in \cite{Sgi,KT1}
by requireing the overlap conditions for the
B-picture variables:
\begin{eqnarray}
 \label{Boverlap}
&& \bra V_3 |\left(\XB^{i(r+1)}(\sigma) 
- \XB^{i(r)}(\pi-\sigma)\right) = 0, 
\quad ( 0 \le \sigma \le \frac{\pi}{2} ) \nn
&& \bra V_3 |\left(\hat{P}_i^{(r+1)}(\sigma) 
+ \hat{P}_i^{(r)}(\pi-\sigma)\right) = 0.
\qquad  ( 0 \le \sigma \le \frac{\pi}{2} )
\end{eqnarray}
Note that this 3-string vertex is the same to
the original one (\ref{overlap}), i.e., the 
one with no background B-field.
In \cite{Sgi,KT1}, it was also shown that
the 3-string vertex (\ref{VhV})
can be rewritten into the following form:
\begin{eqnarray}
 \label{NCV3}
\bra V_3 |
=
\bra \hat{V}_3 | 
\exp
\left(
-\frac{i}{2} \sum_{r=1}^3
\theta^{ij}
 \hat{p}_{i 0}^{(r)} \hat{p}_{j 0}^{(r+1)}
\right) .
\end{eqnarray}


The kinetic term
of the open string field theory
in the B-field background is given by the
B-field dependent BRST charge $\hat{Q}(B)$:
\begin{eqnarray}
 \label{hQ}
\hat{Q}(B) = 
\sum_{n=-\infty}^{\infty}
: c_n \left(\hat{L}_{-n}^X + \frac{1}{2} L_{-n}^{gh}\right) : \,\, ,
\end{eqnarray}
where the matter part of the 
Virasoro operator is given by
\begin{eqnarray}
 \label{hL}
\hat{L}_n^X = 
\frac{1}{2} G_{ij} \sum_{\ell=0}^\infty 
:\hat{\alpha}_{n-\ell}^i \hat{\alpha}_\ell^j : \,\,.
\end{eqnarray}
Here, $:\, \,:$ denotes the normal ordering 
with respect to the operators
\begin{eqnarray}
 \label{halpha}
\hat{\alpha}_n^i = 
\frac{1}{\sqrt{4\pi\alpha'}}
(2\pi\alpha' G^{ij} \hat{p}_{jn} - i n \hat{x}_n^i). 
\end{eqnarray}
The open string field action in the B-field background
is given by
\begin{eqnarray}
 \label{SB}
{\bf S}(B) = 
\frac{1}{2} 
\left\langle R(1,2)\right| 
| \Psi \rangle_1
\left(
\hat{Q}(B)^{(2)} | \Psi \rangle_2
\right)
+\frac{g}{3}\, \,  {}_{123}\left\langle V_3\right|  
| \Psi \rangle_1
| \Psi \rangle_2
| \Psi \rangle_3  ,
\end{eqnarray}
where, as I have explained, the 3-string vertex
$\bra V_3 |$ is the
same to the one without B-field.
This is expected from the background independence
of the 3-string vertex \cite{bgsen,sonoda,bgSZ,BCFTD}.
The information of the background only
appears in the BRST charge $\hat{Q}(B)$.


The algebra of the original coordinates
$x^i_0$ is commutative. 
On the other hand,
one may regard
$\hat{p}_{i0}$, which commutes with hamiltonian 
$\hat{H}(B)$, 
as target space momentum.
Then, the
canonical conjugate variable
$\hat{x}^i_0$ can be 
regarded as target space ``coordinate'',
and multiplication between coordinates
are given by a
non-commutative Moyal type product, as one can observe
from eq.(\ref{NCV3}).
With this ``definition'' of the coordinates,
the hamiltonian $\hat{H}(B)$ determines
the subalgebra
of open string field algebra
which is regarded as algebra of
target space coordinates.
Notice that under the B-field background,
there is no intuitive notion for
target space coordinates before one defines them.

To quantize the worldsheet theory, one should also
choose Hilbert space.
It turns out
$|\bra \hat{0} | e^N | \hat{0} \ket|^2 = 0$ and
$|\bra \hat{0} | e^M | \hat{0} \ket|^2 = 0$.
Here, $| \hat{0} \ket$ is a Fock vacuum for
the annihilation/creation operators (\ref{hcre}).
Therefore, $U_{BF}=e^N e^M$ does not
 act within a single Hilbert space.
But one can use linear map
(\ref{x}), (\ref{p}) or the inverse map (\ref{hx}), (\ref{hp})
to describe F-picture operators in terms B-picture operators
or vice versa.
In open string field algebra,
it is not precisely known what set of states
should be allowed 
(see \cite{Singular} for a recent discussion).
It seems that
one should allow transformations
like $U_{BF}$ to treat different
backgrounds in
open string field algebra.

I close this section by mentioning that
an interpolating picture 
between B-picture and F-picture might 
have been obtained
if I had
partially integrated the second term in (\ref{BWS})
\cite{NCSW,Pioline,Wilson}.


\section{Equivalence Of B-field
And Non-commuting Chan-Paton Matrices
In Open String Field Theory}\label{secD}

There is an open string worldsheet theory
which is equivalent to the open string
theory in B-field background
discussed in the previous section \cite{Ish}.
This is an open string theory
on infinitely many D-branes
with non-commuting Chan-Paton matrices.
I will call this description
D-picture.
More precise description is given in the following.
\subsection{Map Between B-picture and D-picture}


The worldsheet action of what I call
D-picture is given by 
\begin{eqnarray}
 \label{SDgene}
S &=&
-\frac{1}{4\pi\alpha'}
\int\!d\tau \int_0^{\pi}\!d\sigma
\left\{
g_{ij}\eta^{\alpha\beta}
\pa_\alpha X^i \pa_\beta X^j
-
\bar{b}_{ij} \epsilon^{\alpha\beta}
\pa_\alpha X^i \pa_\beta X^j
\right\} \nn
&&-
\frac{1}{2\pi\alpha'}\int\!d\tau
(g_{ij}{X^i}'(\pi)-\bar{b}_{ij}{\dot{X}^i(\pi)} )y^j_{\pi}
+
\frac{1}{2\pi\alpha'}\int\!d\tau
(g_{ij}{X^i}'(0)-\bar{b}_{ij}{\dot{X}^i(0)}) y^j_{0} \nn
&&
-\frac{1}{4\pi\alpha'}\int\,d\tau
\left(\frac{\theta}{2\pi\alpha'}
\right)^{-1}\!\!\!\!\!\! {}_{ij}
(
y_\pi^i \dot{y}_\pi^j
-
y_0^i \dot{y}_0^j),
\end{eqnarray}
where $\bar{b}_{ij}$ is given by
\begin{eqnarray}
 \label{bbar}
\bar{b}_{ij} = b_{ij} -  
\left(
\frac{\theta}{2\pi\alpha'}
\right)^{-1}
\!\!\!\!\!\!\!{}_{ij}.
\end{eqnarray}
In the above
the inverse of $\theta^{ij}$
is considered in the appropliate 
subspace of the target space,
i.e. if the rank of $\theta^{ij}$
is $24$, $\theta^{-1}_{ij}$ is the inverse
in the $24$-dimensional subspace. 
The action (\ref{SDgene}) can be read off
from 
the corresponding boundary state
in closed string theory 
\cite{Ish}\cite{Wilson,OkuB}.
In the boundary state, 
the quantum mechanical degrees of freedom
$y^i_0$ and $y^j_\pi$ arise as
path integral counterparts
of Chan-Paton matrices.
Here, they will become 
infinite dimensional Chan-Paton matrices
after the canonical quantization.
As in the previous section, I introduce
cutoff $\e$ to separate
interactions from
boundary conditions.
Then, the equations of motion are given by
\begin{eqnarray}
g_{ij} (\ddot{\XD}^j(\tau,\sigma)-{\XD^j}''(\tau,\sigma)  )
- g_{ij} (y_\pi^j
{\delta^c}'
(\pi-\epsilon,\sigma)-y_0^j{\delta^c}'(\epsilon,\sigma)  ) \quad &&\nn
+\bar{b}_{ij}
\left( 
\dot{y}_\pi^j 
      \delta^s(\pi-\epsilon,\sigma)
-\dot{y}_0^j 
      \delta^s(\epsilon,\sigma)  
\right)
&=& 0 ,  
\label{eqDxg}\\
g_{ij} {\XD^j}'(\tau,\pi-\epsilon) 
+\bar{b}_{ij}  \dot{\XD^j}(\tau,\pi-\epsilon)
+ \left(
\frac{\theta}{2\pi\alpha'}
\right)^{-1}\!\!\!\!\!\!\!{}_{ij} \dot{y}_\pi^j &=& 0, 
\label{eqDypig}\\
g_{ij} {\XD^j}'(\tau,\epsilon) 
+\bar{b}_{ij}  \dot{\XD^j}(\tau,\epsilon)
+\left(
\frac{\theta}{2\pi\alpha'}
\right)^{-1}\!\!\!\!\!\!\!{}_{ij}\dot{y}_0^j &=& 0,
\label{eqDy0g}
\end{eqnarray}
where the suffix $D$
indicates that
$\XD^i(\tau,\sigma)$ satisfies the
Dirichlet boundary conditions. 
In the above, 
the delta function $\delta^s(\sigma,\sigma')$ for 
functions with Dirichlet
boundary conditions 
is given by
\begin{eqnarray}
 \label{deltas}
\delta^s(\sigma,\sigma')
= 
\sum_{n=1}^\infty
\si_n(\sigma) \si_n (\sigma').
\end{eqnarray}
As in the previous section,
I move to Hamiltonian formalism.
$\XD^i(\sigma)$
can be expanded as
\begin{eqnarray}
 \label{gXmod}
\XD^i(\sigma)
=
\sum_{n=1}^\infty
\tilde{x}^i_n \si_n (\sigma).
\end{eqnarray}
The canonical momentum of $\XD^i(\sigma)$
is given by
\begin{eqnarray}
 \label{gPD}
\tilde{P}_i(\sigma)
&=&
\frac{1}{2\pi\alpha'}
\left(
g_{ij} \dot{\XD}^j(\sigma) 
+
\bar{b}_{ij} {\XD^j}'(\sigma) 
\right)
+
\frac{1}{2\pi\alpha'}
\bar{b}_{ij}
 \left( 
y^j_\pi 
   \delta^s(\pi-\e,\sigma)
 - y^j_0 
\delta^s(\e,\sigma)
 \right).
\end{eqnarray}
$\tilde{P}_i(\sigma)$ can be expanded as
\begin{eqnarray}
 \label{PD}
\tilde{P}_i(\sigma)
=
\sum_{n=0}^\infty
\tilde{\tilde{p}}_{in} \si_n(\sigma) 
+
\bar{b}_{ij} \sum_{n=0}^\infty
n \tilde{x}_n^j
\co_n(\sigma),
\end{eqnarray}
where
\begin{eqnarray}
 \label{dtildp}
\tilde{\tilde{p}}_{in}
=
\frac{1}{2\pi\alpha'}
\left(
g_{ij} \dot{\tilde{x}}_n^j
+
\bar{b}_{ij} 
 \left(
y^j_\pi 
   \si_n (\pi-\e)
 -y^j_0 
   \si_n (\e)
 \right)
\right).
\end{eqnarray}
This expansion can be obtained
in a similar manner that was done in \cite{KT1}
for B-picture to obtain the expansions
(\ref{XhX})$\sim$(\ref{hPmod}).
However, it will turn out
to be more convenient to use the
variable $\tilde{p}_{in}$ defined by
\begin{eqnarray}
 \label{tp}
\tilde{p}_{in}
&=&
\int_0^\pi d\sigma
\frac{1}{2\pi\alpha'}
\tilde{P}_i(\sigma)
\si_n(\sigma) 
\nn
&=&
\tilde{\tilde{p}}_{in} +
\frac{1}{2\pi\alpha'}
\bar{b}_{ij} 
\sum_{\ell = 0}^\infty
\tilde{x}_\ell^j \ell A_{\ell n} n,
\end{eqnarray}
to treat the subtle prescription for
summation over mode indices
similar to the one appeared in (\ref{boundary}).
An underling reason is that
the variable in B-picture corresponding to
(\ref{gPD}) is
\begin{eqnarray}
 \label{BhX'}
\frac{1}{2\pi\alpha'}
\left(
g_{ij}\dot{\XB}^j(\sigma)
+
\bar{b}_{ij} {\XB^j}'(\sigma)
\right)
=
\theta^{-1}_{ij}
\hat{X}^j{}'(\sigma),
\end{eqnarray}
and it is expanded by $\si_n(\sigma)$.
Since $\sum_{n=1}^\infty \tilde{p}_{in} \si_n(\sigma)$
is a part of $\tilde{P}_{i}(\sigma)$ which can be
expanded by $\si_n(\sigma)$,
the matching between them contains no subtlety at boundaries.
Notice that
the map between  $\tilde{\tilde{p}}_{in}$ and $\tilde{p}_{in}$
can be written in the form of unitary
transformation
\begin{eqnarray}
 \label{ttptp}
\tilde{p}_{in} =
e^{\tilde{N}_{\bar{b}}}
\tilde{\tilde{p}}_{in} 
e^{-\tilde{N}_{\bar{b}}},
\end{eqnarray}
where
\begin{eqnarray}
 \label{Nb}
\tilde{N}_{\bar{b}} &=& 
\frac{i}{4\pi\alpha'} \bar{b}_{ij}
\sum_{\ell= 1}^\infty \sum_{m=1}^\infty
\tilde{x}_\ell^i \ell A_{\ell m} m  \tilde{x}_m^j 
\nn
&=&
-\frac{i}{4\pi\alpha'}
\int_0^\pi
d\sigma\,
\bar{b}_{ij}
\XD^i(\sigma){\XD^j}'(\sigma).
\end{eqnarray}
The map (\ref{ttptp}) can be regarded as
a map from the variables without background B-field
to those with constant 
background B-field $\bar{b}_{ij}$
in open string theory
on lower dimensional
D-branes.
This is another reason 
that the variable $\tilde{p}_{in}$ is convenient.

The canonical commutation relations between modes are
given by
\begin{eqnarray}
 \label{commD}
&&[\tilde{x}^i_n, \tilde{p}_{jm}]
=
i \delta_{nm}, \\
\label{commy}
&&[y^i_\pi, y^j_\pi] = - i \theta^{ij}, \quad
[y_0^i,y_0^j] = i \theta^{ij},
\end{eqnarray}
others give zero.
The Hamiltonian of this system
is given by
\begin{eqnarray}
 \label{HD}
\HD
=
\frac{1}{4\pi\alpha'}
g_{ij}
\sum_{n=1}^\infty
\left(
\dot{\tilde{x}}^i_n\dot{\tilde{x}}^j_n
+
n^2 \tilde{x}^i_n\tilde{x}^j_n
\right)
-
\frac{1}{2\pi\alpha'}
g_{ij}
\left(
y^j_\pi \tilde{x}^i_n n \co_n(\pi-\e)
-
y^j_0  \tilde{x}^i_n n \co_n(\e)
\right),
\end{eqnarray}
where $\tilde{x}^i_n$
is solved in terms of
$\tilde{x}^i_n$
and $\tilde{p}_{i n}$ by (\ref{dtildp}) and (\ref{tp}):
\begin{eqnarray}
 \label{geqxD}
\dot{\tilde{x}}^i_n
&=& 
 2\pi\alpha' g^{ij} \tilde{p}_{j n} 
-
{(g^{-1}\bar{b})^i}_{j}
 \left(
\sum_{\ell=0}^\infty \tilde{x}^j_\ell \ell A_{\ell n} n
-
  \left(
y^j_\pi
   \si_n(\pi-\e)
-y^j_0 
   \si_n(\e)
 \right)
\right).
\end{eqnarray}
This is also an equation of motion of $\tilde{x}^i_n$.
Other equations of motion become
\begin{eqnarray}
 \label{geqpD}
\dot{\tilde{p}}_{in} 
&=& i[\HD , \tilde{p}_{in}] \nn
&=& -\frac{1}{2\pi\alpha'} 
\left( g-\bar{b}g^{-1}\bar{b} \right)_{ij} 
n
\left(
n \tilde{x}^j_n + 
 y^j_\pi \co_n (\pi-\e)
-y^j_0  \co_n (\e)
\right)  \nn
&&+
{(\bar{b}g^{-1})_i}^j 
\sum_{\ell=1}^\infty
\tilde{p}_{j\ell} \ell A_{\ell n} n  , \label{geqypi}\\
\dot{y}_\pi^i 
&=& i [\HD, y^i_\pi]
= - \frac{\theta^{ij}}{2\pi\alpha'}
\left(
 g_{jk} {\XD^k}'(\pi-\e) + 
\bar{b}_{jk} \dot{\XD}^k(\pi-\e)
\right) , \\
 \label{geqy0}
\dot{y}_0^i 
&=& i [\HD, y^i_0]
= - \frac{\theta^{ij}}{2\pi\alpha'}
\left(
 g_{jk} {\XD^k}'(\e) + 
\bar{b}_{jk} \dot{\XD}^k(\e)
\right).
\end{eqnarray}
%
%
Now, I take the
$\e \rightarrow 0$ limit
of the equations of motion
(\ref{geqpD})$\sim$(\ref{geqy0})
with 
the conditions
\begin{eqnarray}
 \label{limitD}
&& \lim_{\e \rightarrow 0}
\XD^i (\pi-\e) = \XB^i(\pi) \ne \XD^i (\pi), \nn
&& \lim_{\e \rightarrow 0}
\XD^i (\e) = \XB^i(0) \ne \XD^i (0) .
\end{eqnarray}
This prescription can be understood
from the corresponding
boundary state in closed string field theory
\cite{Ish}\cite{Wilson,OkuB}.
Then, I obtain
\begin{eqnarray}
 \label{limgeqxD}
\dot{\tilde{x}}^i_n
&=&  2\pi\alpha' g^{ij} \tilde{p}_{j n}
-(g^{-1}\bar{b}){{}^i}_j 
\sum_{\ell=0}^{\infty} 
\tilde{x}_\ell^j
\ell A_{\ell n} n, \\
 \label{limgeqpD}
\dot{\tilde{p}}_{in} 
&=& -\frac{1}{2\pi\alpha'} 
\left( g-\bar{b}g^{-1}\bar{b} \right)_{ij} 
n 
\left(
n\tilde{x}^j_n 
+
 y^j_\pi  \co_n (\pi)
-y^j_0  \co_n (0)
\right) 
\nn
&&+
{(\bar{b}g^{-1})_i}^j 
\sum_{\ell=1}^\infty
\tilde{p}_{j\ell} \ell A_{\ell n} n , \\
\label{limgeqypi}
\dot{y}_\pi^i 
&=& - \frac{\theta^{ij}}{2\pi\alpha'}
\left(
 g_{jk} {\XB^k}'(\pi) + 
\bar{b}_{jk} \dot{\XB}^k(\pi)
\right), \\
 \label{limgeqy0}
\dot{y}_0^i 
&=& - \frac{\theta^{ij}}{2\pi\alpha'}
\left(
 g_{jk} {\XB^k}'(0) + 
\bar{b}_{jk} \dot{\XB}^k(0)
\right).
\end{eqnarray}
To obtain the map between B-picture and D-picture,
one equates $\XD^i(\sigma)=\XB^i(\sigma)$
in the interval ($0 < \sigma < \pi$).
In this interval, one can
expand $\co_n(\sigma)$ by $\si_m(\sigma)$ and
the coefficients of $\si_m(\sigma)$ in the 
both side should match and one obtains
the expression of $\tilde{x}_n^i$ in terms
of B-picture variables.
Then, the expression of 
$\tilde{p}_{in}$, $y_\pi^i$ and $y_0^i$
can be obtained
from (\ref{limgeqxD})$\sim(\ref{limgeqy0})$.
The result is as follows:
\begin{eqnarray}
\tilde{x}^i_n
&=&
\sum_{\ell=0}^\infty
\hat{x}^i_\ell A_{\ell n} n + 
\theta^{ij}
\left[
\hat{p}_{j0} \frac{A_n}{n} + \hat{p}_{jn} \frac{1}{n}
\right], \label{gdxnh}\\
\tilde{p}_{in}
&=&
\theta^{-1}_{ij}\hat{x}^j_n n,
 \label{gdpnh} \\
y_\pi^i &=& 
\sum_{n=0}^{\infty} \hat{x}_n^i \co_n(\pi) 
+ \frac{\sqrt{\pi}}{2}\theta^{ij} \hat{p}_{j0}
=
\XB^i(\pi), \label{gypih}\\
y_0^i &=& 
\sum_{n=0}^{\infty} \hat{x}_n^i \co_n(0)
- \frac{\sqrt{\pi}}{2} \theta^{ij}\hat{p}_{j0}
= \XB^i(0). \label{gy0h}
\end{eqnarray}
Using the commutation relations in
B-picture (\ref{commB}),
one can check that
(\ref{gdxnh})$\sim$(\ref{gy0h})
satisfy the commutation relations 
in D-picture (\ref{commD})
and (\ref{commy}).
One can also check 
that the equations of motion for
the variables in B-picture 
lead to the equations of motion for D-picture
(\ref{limgeqxD})$\sim$(\ref{limgeqy0}).
This means that
both the equations of motion in B-picture and
those in D-picture are given by
the same Hamiltonian $\hat{H}(B)$.
Thus, these two descriptions are equivalent.

The inverse map is given by
\begin{eqnarray}
\hat{x}^i_n
&=&
\theta^{ij} \frac{1}{n} \tilde{p}_{jn},
\quad (n \ne 0) 
\label{ghxnd}
\\
\hat{x}_0^i
&=&
\frac{\sqrt{\pi}}{2}
\left(
y_\pi^i + y_0^i
\right)
-\frac{\sqrt{\pi}}{2} \theta^{ij}
\sum_{\ell=1}^\infty
\tilde{p}_{j \ell}
\frac{\co_\ell(\pi) + \co_\ell(0)}{\ell}, 
\label{ghx0d}\\
\hat{p}_{in}
&=&
- \sum_{\ell = 1}^\infty
\tilde{p}_{i \ell} \ell A_{\ell n}
+
\theta^{-1}_{ij}
\left(
n \tilde{x}_n^j +
  \left(
 y_\pi^j   \co_n(\pi) -
 y_0^j   \co_n(0)  
  \right) 
\right).
\label{ghpd}
\end{eqnarray}
Notice that the swallow of
Chan-Paton matrices in (\ref{ghpd})
is similar to the phenomenon appeared in \cite{AL}.
The similarity  becomes clearer 
if one rewrites the
Hamiltonian $\hat{H}(B)$
in terms of D-picture variables.\footnote{
I learned this resemblance
from N. Ishibashi.}

One can write the map in the form of 
unitary transformation
\begin{eqnarray}
 \tilde{U}_{\scriptscriptstyle BD} \hat{x}^i_0 
\tilde{U}_{\scriptscriptstyle BD}^{-1} 
&=&
\frac{\sqrt{\pi}}{2} 
\left(y^i_\pi+ y^i_0 \right),
\label{tBDx0} \\
\tilde{U}_{\scriptscriptstyle BD} \hat{p}_{i0} 
\tilde{U}_{\scriptscriptstyle BD}^{-1}
&=&
\frac{1}{\sqrt{\pi}}\theta^{-1}_{ij}
\left(y^j_\pi- y^j_0 \right),
\label{tBDp0} \\
\tilde{U}_{\scriptscriptstyle BD} \hat{x}_n^i 
\tilde{U}_{\scriptscriptstyle BD}^{-1}
&=&\theta^{ij}  \frac{1}{n} \tilde{p}_{jn}, \quad (n \ne 0)
\label{tBDxn} \\
\tilde{U}_{\scriptscriptstyle BD} \hat{p}_{i n} 
\tilde{U}_{\scriptscriptstyle BD}^{-1}
&=&  \theta^{-1}_{ij} \tilde{x}_n^j n, \quad (n \ne 0)
\label{tBDpn}
\end{eqnarray}
where
\begin{eqnarray}
\tilde{U}_{BD} = e^{C_1}e^{C_2}e^{C_3}e^{C_4},
\end{eqnarray}
\begin{eqnarray}
C_1 &=& i \frac{\sqrt{\pi}}{2} \hat{p}_{i0}
\sum_{n=1}^\infty
\hat{x}^i_n
\left(
\co_n(\pi) + \co_n(0)
\right),
\label{C1} \\
C_2&=&
-\frac{i}{\sqrt{\pi}}
\theta^{-1}_{ij}
\hat{x}^i_{0}
\sum_{n=1}^\infty
\hat{x}^j_n
\left(
\co_n(\pi) - \co_n(0)
\right),
\label{C2} \\
C_3 &=&
-\frac{i}{2}
\theta^{-1}_{ij}
\sum_{\ell=1}^\infty \hat{x}_\ell^i \co_\ell(\pi)
\sum_{m=1}^\infty \hat{x}^j_m \co_m(0),
\label{C3}\\
C_4 &=&
\frac{i}{2}
\theta^{-1}_{ij}
\sum_{\ell = 1}^\infty
\sum_{m=1}^\infty
\hat{x}^i_\ell A_{\ell m} \frac{\ell^2+m^2}{2}
\hat{x}^j_m.
\label{C4}
\end{eqnarray}
From (\ref{tBDx0})$\sim$(\ref{tBDpn}),
it is easily understood that one can 
further transform them by
Lorenz rotation,
rotation in phase space and scale transformation
to obtain $U_{\scriptscriptstyle BD}$ which satisfies 
\begin{eqnarray}
 U_{\scriptscriptstyle BD} \hat{x}^i_0 U_{\scriptscriptstyle BD}^{-1} 
&=&
\frac{\sqrt{\pi}}{2} 
\left(y^i_\pi+ y^i_0 \right).
\label{BDx0} \\
U_{\scriptscriptstyle BD} \hat{p}_{i0} U_{\scriptscriptstyle BD}^{-1}
&=&
\frac{1}{\sqrt{\pi}}\theta^{-1}_{ij}
\left(y^j_\pi- y^j_0 \right),
\label{BDp0} \\
U_{\scriptscriptstyle BD} \hat{x}_n^i U_{\scriptscriptstyle BD}^{-1}
&=& \tilde{x}^i_n, \quad (n \ne 0)
\label{BDxn} \\
U_{\scriptscriptstyle BD} \hat{p}_{i n} U_{\scriptscriptstyle BD}^{-1}
&=&  \tilde{p}_{in}. \quad (n \ne 0)
\label{BDpn}
\end{eqnarray}

Notice that from 
(\ref{gdxnh})$\sim$(\ref{gy0h}), 
the 3-string vertex
$\bra V_3 | = \bra \hat{V}_3 |e^M$ satisfies
the following overlap conditions:
\begin{eqnarray} 
 \label{Doverlap}
&&\bra {V}_3 | 
\left(
\XD^{i(r+1)}(\sigma) - \XD^{i(r)}(\pi-\sigma)
\right)
=0, \quad (0 < \sigma \le \frac{\pi}{2}) \nn
&& \bra {V}_3 | 
\left(
\tilde{P}^{(r+1)}_i (\sigma) + \tilde{P}^{(r)}_i(\pi-\sigma)
\right)
=0, \qquad (0 < \sigma \le \frac{\pi}{2}) \nn
&& \bra {V}_3 | 
\left(y_0^{i(r+1)} - y_\pi^{i(r)} \right) = 0 .
\end{eqnarray}
These are natural overlap conditions for
open string field theory on lower dimensional
D-branes.

\subsection{3-string Vertex %
In D-picture}

In this subsection I study
another 3-string vertex which is 
naturally obtained in D-picture
(see \cite{Klu} for related issues).
The 3-string vertex is divided into
the overlap vertex
for $\XD^i(\sigma)$ part 
and Chan-Paton matrix part.

I first construct the
Chan-Paton matrix multiplication part.
In the following, I assume the rank of
the matrix $\theta^{ij}$ is $24$. 
I set $\theta^{0i}=\theta^{1i}=0$,
and set $\theta^{ij}=0$ except
$\theta^{2i\, 2i+1} =-\theta^{2i+1\, 2i}$,
by an appropriate rotation in the target space.
Also I only consider the case where
$\theta^{2i\, 2i+1} > 0$.
Generalization to other
cases is straightforward.

First I define creation and annihilation
operators in Chan-Paton algebra 
as follows.
\begin{eqnarray}
 \label{zs}
z_0^{i} = 
 \sqrt{\frac{1}{2\theta^{2i\,2i+1}}} 
 \left(
   y^{2i}_0 + i y_0^{2i+1}
 \right), \quad
z^{i \dagger}_0 = 
 \sqrt{\frac{1}{2\theta^{2i\,2i+1}}} 
 \left(
  y^{2i}_0 - i y_0^{2i+1}
 \right), \nn
z_\pi^{i} = 
 \sqrt{\frac{1}{2\theta^{2i\,2i+1}}} 
 \left(
  y^{2i}_\pi - i y_\pi^{2i+1}
 \right), \quad
z^{i \dagger}_\pi =  
 \sqrt{\frac{1}{2\theta^{2i\,2i+1}}} 
  \left(
    y^{2i}_\pi + i y_\pi^{2i+1}
\right).
\end{eqnarray}
Their commutation relations are given by
\begin{eqnarray}
 \label{commz}
[z_0^{i}, z^{j \dagger}_0 ] = \delta^{ij}, 
\quad
[z_\pi^{i}, z^{j \dagger}_\pi ] = \delta^{ij}, 
\end{eqnarray}
and otherwise zero.
These operators act on the
Hilbert space whose basis is
labeled by
${\bf n} = (n_1, n_2, \cdots, n_{12})$:
\begin{eqnarray}
 | {\bf n}\ket_0   
&=& \prod_{i=1}^{12}
\frac{1}{\sqrt{n_i!}} (z_0^{i\dagger})^{n_i}   |0\ket_0 ,
\\
 | {\bf n}\ket_\pi 
&=& \prod_{i=1}^{12}
\frac{1}{\sqrt{n_i!}} (z_\pi^{i\dagger})^{n_i} |0\ket_\pi,
\end{eqnarray}
where $|0\ket_0$ and $|0\ket_\pi$ are
defined by the conditions
\begin{eqnarray}
 z_0^i   |0\ket_0 = 0 , \quad
 z_\pi^i |0\ket_\pi = 0 .
\end{eqnarray}
Arbitrary open string field in D-picture
has Chan-Paton indices
\begin{eqnarray}
 \label{CPind}
|{\bf n} \ket_0
|{\bf m} \ket_\pi .
\end{eqnarray}
$|{\bf n} \ket_0|{\bf m} \ket_\pi$
itself is regarded as
Chan-Paton matrix.
The matrix multiplication is
represented by 
$\bra v_3^{CP} |$
defined as follows:
\begin{eqnarray} 
 \label{CPV3}
\bra v_3^{CP} |
&=&
\prod_{r=1}^3
\sum_{{\bf n}}
{}_{\,\,\, \pi}^{(r)}
\bra {\bf n} |\,\,{}^{(r+1)}_{\quad \,\,\,0}
\bra {\bf n} | \nn
&=&
\prod_{r=1}^3 
{}_{\,\,\, \pi}^{(r)}
\bra 0 |\,\,{}^{(r+1)}_{\quad \,\,\,0}\bra 0 |
\exp
\left(
\sum_{i=1}^{12}
 z^{i(r)}_\pi z_0^{i(r+1)} 
\right).
\end{eqnarray}
The first line is a definition
of the usual matrix multiplication.
The form in the
second line is a simpler analog of
the vertex which appeared in the
matrix representation of the 3-string vertex
\cite{MatSFT}.
Precisely  speaking,
here there appear three kinds of Chan-Paton matrices:
those acting on the vector $|{\bf n}\ket_0$,
those acting on $|{\bf n}\ket_\pi$,
and those whose basis is given by
(\ref{CPind}).
But those differences are not essential.
The basis of the first type is given by
$| {\bf n} \ket_0 {}_0\!\bra {\bf m} |$ and
the basis for the second type is given by
$| {\bf n} \ket_\pi {}_\pi\!\bra {\bf m} |$.
One can isomorphically convert them into the
third type by multiplying them to the identity
matrix in the third type.
\begin{eqnarray}
 \label{id}
| {1} \ket 
= 
\sum_{{\bf n},{\bf m}} 
\delta_{{\bf n}{\bf m}}
| {\bf n} \ket_0 | {\bf m} \ket_\pi ,
\end{eqnarray}
\begin{eqnarray}
 \label{isoCP}
| {\bf n} \ket_0 {}_0\!\bra {\bf m} | | {1} \ket 
=| {\bf n} \ket_0 | {\bf m} \ket_\pi,  \quad
| {\bf n} \ket_\pi {}_\pi\!\bra {\bf m} | | {1} \ket 
= | {\bf m} \ket_0 | {\bf n} \ket_\pi .
\end{eqnarray}
As an example, let us consider a map
\begin{eqnarray}
 \label{incl}
z_0^{i \dagger} \mapsto 
 z_{0}^{i \dagger} | {1} \ket =
 z_{\pi}^{i}  | {1} \ket , \quad
z_0^i \mapsto 
 z_{0}^{i} | {1} \ket =
 z_{\pi}^{i \dagger}  | {1} \ket .
\end{eqnarray}
Then, the algebra (\ref{commz}) is
isomorphically mapped to 
\begin{eqnarray}
 \label{SFmat}
 [z_{0}^{i \dagger} | {1} \ket ,
 z_{0}^{j \dagger} | {1} \ket ]_{CP}
= \delta^{ij} | {1} \ket,
\end{eqnarray}
where the subscript $CP$ denotes that the
multiplication is defined by (\ref{CPV3}).

The 3-string vertex in D-picture can be
constructed as
\begin{eqnarray}
 \label{V3D}
\bra V_3^D | =
\bra V_3^{X_D} |
\bra v_3^{CP} | ,
\end{eqnarray}
where $\bra V_3^{X_D} |$ is defined by the overlap
conditions
\begin{eqnarray}
 \label{overlapD}
&&\bra V_3^{X_D} | 
\left(
\XD^{i(r+1)}(\sigma) - \XD^{i(r)}(\pi-\sigma)
\right)
=0, \quad (0 < \sigma \le \frac{\pi}{2}) \nn
&& \bra V_3^{X_D} | 
\left(
\tilde{P}^{(r+1)}_i (\sigma) + \tilde{P}^{(r)}_i(\pi-\sigma)
\right)
=0. \qquad (0 < \sigma \le \frac{\pi}{2}) 
\end{eqnarray}
Since both $\bra V_3|$ and $\bra V_3^D |$
satisfy the same overlap condition,
they are identical
up to normalization.
It would be interesting to
check this
by a direct calculation 
using explicit expressions for Neumann coefficients 
of the 3-string vertex
and the map between different pictures.

\subsection{Chan-Paton Algebra In Open String Field Algebra}

Since the map between
F-picture,
B-picture and D-picture has been obtained,
now 
I can identify
the 
Chan-Paton matrix subalgebra
in the original 
open string field algebra.
Let us extract the Chan-Paton
algebra in open string field algebra
whose star product is defined by (\ref{star}),
without using
the expression (\ref{V3D}).
As can be expected 
from (\ref{isoCP}),
to extract Chan-Paton subalgebra 
from open string field algebra,
what I need to do is to just multiply
$\XB^{i}(0)$ to the identity state
\begin{eqnarray}
 \label{mapalg}
\XB^{i}(0) \mapsto \XB^i(0) |I\ket.
\end{eqnarray}
This is a map from worldsheet variable
to open string field.
Recall that this kind of map 
from worldsheet operator algebra to 
open string field algebra
often appears
in the cubic open string field theory 
\cite{pure,MatSFT}.
I obtain
\begin{eqnarray}
 \label{SFTalg}
\XB^i(0) |I\ket \star
\XB^j(0) |I\ket
=
\XB^i(0) \XB^j(0) |I\ket ,
\end{eqnarray}
\begin{eqnarray}
 \label{CPalg}
[\XB^{i}(0) |I\ket,
\XB^{j} (0) |I\ket]_\star
=
[\XB^{i}(0),
\XB^{j}(0)] |I\ket
=
i \theta^{ij} |I\ket.
\end{eqnarray}
Here, I have used the relations
\begin{eqnarray}
 \label{hII}
 e^{-M}e^{-N} | I \ket = | I \ket,
\end{eqnarray}
\begin{eqnarray}
 \label{hIoverlap}
\left(\XB^i(\sigma) - \XB^i(\pi-\sigma)\right) | I \ket &=& 0, \\ 
\left(\hat{P}_i(\sigma)+\hat{P}_i(\pi-\sigma) \right) | I \ket &=& 0.
\end{eqnarray}
These relations are obtained in the similar way
that was done for the 3-string vertex $\bra V_3|$.
Thus, the Chan-Paton algebra
(\ref{commy})
is extracted from the open string field algebra.
The map
(\ref{gypih}) and (\ref{gy0h})
confirms that
the algebra (\ref{CPalg})
is identified with the Chan-Paton algebra.
Once the Chan-Paton subalgebra
(\ref{CPalg}) in open string field algebra is
obtained, one can
apply techniques in non-commutative solitons
to obtain exact/approximate solutions.
Notice that in the map
between the different pictures (F,B,D),
no special limit such as 
the so called Seiberg-Witten limit \cite{NCSW}
is assumed.

Let us call the algebra
generated by the string coordinate fields
at the boundaries
as boundary algebra.
What I have shown here is 
that in constant
B-field background,
the boundary algebra {\em is} 
Chan-Paton algebra.
It is easy to show that
using the defining  property of
the 3-string vertex and the identity
field
one can map arbitrary boundary algebra
to open string field algebra
by the map (\ref{mapalg}).
Therefore, 
I expect that
arbitrary boundary algebra
can be interpreted as
Chan-Paton algebra.

\section{Summary And Discussions}\label{summary}
In this article, I have studied
three equivalent
descriptions of open string in constant B-field 
background
in the framework of
Witten's cubic open string field theory.

The equivalence
of B-picture and F-picture
gives the map between the theory with
non-commutative coordinates and
commutative coordinates.
Since F-picture variables
are the original variables, i.e.
the variables without the B-field background,
this means
that the algebra of non-commutative coordinates
is already contained in the
original open string field algebra.

On the other hand,
from the equivalence of
B-picture and D-picture
I obtained the expression of
Chan-Paton matrices
in terms of B-picture variables.
The Chan-Paton matrix in D-picture is
shown to be the string coordinate field
at the boundary in B-picture.
I believe  
in order to find D-branes in open string field theory,
one should not overlook this property, i.e.
the Chan-Paton indices are on the boundaries.
Using the map,
I extracted
the Chan-Paton algebra
as subalgebra in
open string field algebra. 
Since Chan-Paton algebra is
associated with 
lower dimensional D-branes,
this ensures that lower dimensional 
D-branes can be described
as some
field configurations
in Witten's cubic open string field theory.

There are two directions for the
classification of D-branes.
The one is to use
the highest dimensional branes \cite{KM,KWitt}
to classify
lower dimensional D-branes
as topological defects.
The other uses the
lowest dimensional branes \cite{AST}
to construct higher dimensional D-branes.
In this article, 
I started from 
the open string field theory on
space-time filling D-branes,
but I could have started from
open string theory on
infinitely many lower dimensional
D-branes.
Thus, open string field algebra
naturally unifies the two descriptions and 
appears as
the most fundamental framework for the classification of 
D-branes.
In this article, I rewrite
Chan-Paton indices, which are
fundamental variables in D-picture, to
linear combination of variables in another picture.
This strategy may be generalized
and will be a powerful tool to find
D-brane solutions in
Witten's cubic open string field theory.
However, note that this strategy is
applicable to the backgrounds
which can be described by the
same 3-string vertex.
In other words,
open string field algebra
can be used to classify
D-branes which are
contained in that paticular
open string field algebra \cite{Aki}.
It will be important to investigate
to what extent the
3-string vertex can describe
different backgrounds.

The equivalence between
boundary algebra
and Chan-Paton algebra seems
to hold more generally.
It would be interesting to investigate
the relation to the boundary algebra 
\cite{Gera} in
boundary open
string field theory \cite{BOSFT}
along the line of \cite{Nakatsu},
where a proposal for a
concrete definition of the boundary
open string field theory 
was given
as a limit of
continuous deformation from
Witten's cubic open string field theory.

\vskip 10mm
\centerline{\bf Acknowledgments}
\vs{1.5}
\noindent
I would like to express special thanks to 
N. Ishibashi
for collaboration in early stage of this work
as well as successive discussions.
I would also like to thank
F. Sugino for illuminating discussions.
I would like to thank 
A. Futaki, A. Hashimoto, N. Nekrasov,
G. Semenoff and T. Tani for 
questions, suggestions and pleasant conversations. 
Part of this work was done in KEK, and also in the IHES.
I would like to thank my previous colleagues in KEK
for pleasant time.
I am grateful to N. Nekrasov and the IHES for warm hospitality.
I am also grateful 
to the organizers of String Theory in greater Paris,
especially B. Pioline,
for giving me an opportunity
to present some of the results above.
This work was supported in part by
the Pacific Institute for the Mathematical Sciences.

\newpage
\appendix
\section{Formulas %
For The Fourier Coefficients}
%
Ortho-normal basis 
for the functions in $0 \le \sigma \le \pi$ 
with Neumann boundary conditions at the boundaries
$\sigma = 0,\pi$:
\begin{eqnarray}
 \label{defcoa}
&&\co_n(\sigma) = \sqrt{\frac{2}{\pi}} \cos n \sigma,
 \quad (n \ne 0) \nn
&&\co_0(\sigma) = \sqrt{\frac{1}{\pi}} \quad.
\end{eqnarray}
Ortho-normal basis 
for the functions in $0 \le \sigma \le \pi$ 
with Dirichlet boundary conditions at the boundaries
$\sigma = 0,\pi$:
\begin{eqnarray}
 \label{defsa}
\si_n(\sigma) = \sqrt{\frac{2}{\pi}} \sin n \sigma  .
\end{eqnarray}
Definition of the matrix $A_{mn}$:
\begin{eqnarray}
A_{mn} = \frac{1}{2}
\int_0^\pi d\sigma
\int_0^\pi d\sigma'
\epsilon(\sigma-\sigma') \co_m(\sigma) \co_n(\sigma'),
\end{eqnarray}
where $\epsilon(\sigma)$ is the sign function
\begin{eqnarray}
 \label{signa}
\epsilon(\sigma) &=& 1, \quad \,\,\, (\sigma > 0 )\nn
                 &=& 0, \quad \,\,\, (\sigma = 0 )\nn
                 &=& -1. \quad (\sigma < 0 )
\end{eqnarray}
Explicit value of the components:
\begin{eqnarray}
 \label{Amn}
A_{mn} &=&
\frac{1}{m^2-n^2}
\left(
\co_m(\pi)\co_n(\pi)-\co_m(0)\co_n(0)
\right).\quad (m\ne n) \nn
&=& 0. \quad (m=n)
\end{eqnarray}
Formulas:
\begin{eqnarray}
 \label{Acosi}
&&\co_n(\sigma) = \sum_{m=1}^\infty A_{nm} m \si_m(\sigma), 
\quad (0<\sigma<\pi)\\
&&\si_n(\sigma) = \sum_{m=0}^\infty (-n) A_{nm}  \co_m(\sigma),
\quad (0\le\sigma\le\pi)
\end{eqnarray}
\begin{eqnarray}
 \label{fA1}
\sum_{\ell=0}^\infty
A_{n\ell}A_{m\ell}
&=& \frac{1}{nm} \delta_{nm}, 
\qquad \qquad \qquad  \qquad \, (n\ne 0, \, m\ne0)
\nn
&=& -\frac{\sqrt{\pi}}{2m^2} 
 \left(
  \co_m(\pi)+\co_m(0)
 \right), 
\quad (n=0,\, m\ne 0)\nn
&=& -\frac{\pi^2}{12} ,\qquad \qquad \qquad \qquad \qquad (n=m=0)
\end{eqnarray}
\begin{eqnarray}
 \label{fA2}
&&\sum_{\ell=1}^\infty
A_{n\ell}A_{m\ell}{\ell^2} 
= \delta_{nm} .
\end{eqnarray}
\begin{eqnarray}
\sqrt{\frac{1}{\pi}}
\left(
\sigma-\frac{\pi}{2}
\right)
=
-\sum_{\ell =1}^\infty
A_{0\ell} \co_\ell(\sigma). \quad (0 \le \sigma \le \pi)
\end{eqnarray}
\begin{eqnarray}
 \label{aa}
\sum_{\ell=0}^\infty
A_{n \ell }\co_\ell(\pi) 
=- \delta_{n0} \frac{\sqrt{\pi}}{2},\quad 
\sum_{\ell=0}^\infty
A_{n \ell}\co_\ell(0)
 = \delta_{n0} \frac{\sqrt{\pi}}{2}.
\end{eqnarray}
Definition of $A_n$:
\begin{eqnarray}
\frac{1}{n}A_n &=&
\int_0^\pi d\sigma
\sqrt{ \frac{1}{\pi} }
\left(\sigma-\frac{\pi}{2}\right)
\si_n(\sigma) \nn
&=&
-\frac{1}{\sqrt{2}}\frac{1}{n} \left((-)^n+1\right)
= -\frac{\sqrt{\pi}}{2n}
\left( \co_n(\pi) + \co_n(0) \right).
\end{eqnarray}

\section{Formulas For The Target Space Tensors}

\begin{eqnarray}
 \label{AG}
G_{ij} &=& g_{ij} - (b g^{-1} b)_{ij}, \quad
G^{ij} =
\left(
\frac{1}{g+b} g \frac{1}{g-b}
\right)\!{}^{ij}.
\end{eqnarray}

\begin{eqnarray}
 \label{Atheta}
\frac{\theta^{ij}}{2\pi \alpha'}
&=& -  
\left(
\frac{1}{g+b} b \frac{1}{g-b}
\right)\!{}^{ij}.
\end{eqnarray}

\begin{eqnarray}
 \label{Abb}
b_{ij} = \bar{b}_{ij} + 
\left(
\frac{\theta}{2\pi\alpha'}
\right)^{-1}
\!\!\!\!\!\!{}_{ij} \,\, .
\end{eqnarray}

\begin{eqnarray}
 \label{Agth}
{\left(bG^{-1} \right)_i}^j
= -
\left(
\frac{g\theta}{2\pi\alpha'}
\right)\!{{}_{i}}^j ,\quad
\left(
\frac{\bar{b}\theta}{2\pi\alpha'}
\right)\!{{}_{i}}^j
=
-\left(
gG^{-1}
\right)\!{{}_i}^j.
\end{eqnarray}
\newpage

\end{document}